\documentclass[aps, prl, twocolumn, showpacs, preprintnumbers,
amssymb, floatfix]{revtex4-1}

\usepackage{amsmath}
\usepackage{amsfonts}
\usepackage{amsbsy}
\usepackage{xcolor}
\usepackage{soul}
\usepackage{graphicx}

\newcommand{\spvec}[1]{\ensuremath{\mathbf{#1}}}
\newcommand{\unitvec}[1]{\ensuremath{\mathbf{\hat{#1}}}}

\newcommand{\colvec}[1]{\ensuremath{\mathrm{#1}}}

\renewcommand{\(}{\left(}
\renewcommand{\)}{\right)}

\newcommand{\commentout}[1]{{}}

\newcommand{\beq}{\begin{equation}}
\newcommand{\eeq}{\end{equation}}

\renewcommand{\>}{\rangle}

\newcommand{\sje}{\mathrm{e}}
\newcommand{\sjm}{\mathrm{m}}
\newcommand{\sjo}{\mathrm{o}}

\newcommand{\cbE}{\boldsymbol{\mathbf{\cal E}}}
\newcommand{\cbH}{\boldsymbol{\mathbf{\cal H}}}

\begin{document}
\title{Many-body subradiant excitations in metamaterial arrays: Experiment and theory}
\author{Stewart D. Jenkins and Janne Ruostekoski}
\affiliation{Mathematical Sciences and Centre for Photonic
	Metamaterials, University of Southampton,
	Southampton SO17 1BJ, United Kingdom}
\author{Nikitas Papasimakis$^1$, Salvatore Savo$^{1,2}$, and Nikolay I. Zheludev$^{1,3}$}
\affiliation{$^1$Optoelectronics Research Centre and Centre for Photonic Metamaterials,
	University of Southampton Southampton SO17 1BJ, United Kingdom\\
$^2$TetraScience Inc.,
114 Western Ave,
Boston, MA, 02134\\
$^3$Centre for Disruptive Photonic Technologies, \\School of Physical and Mathematical Sciences and The Photonics Institute, Nanyang Technological University, Singapore 637378, Singapore
}

\begin{abstract}
Subradiant excitations, originally predicted by Dicke, have posed a long-standing challenge in physics owing to their weak radiative coupling to environment.
Here we engineer massive subradiance in planar metamaterial arrays as a spatially extended eigenmode comprising over 1000 metamolecules.
By comparing the near- and far-field response in large-scale numerical simulations with those in
experimental observations we identify correlated multimetamolecule subradiant states
that dominate the total excitation energy.
We show that spatially extended many-body subradiance can also exist in plasmonic metamaterial arrays at optical frequencies.
\end{abstract}
\date{\today}

\maketitle

The classic example of neutrons and magnetic dipole radiation by Dicke~\cite{Dicke1954} over 60 years ago describes the collective superradiant and subradiant response of emitters at high density. Superradiance, where the emission is enhanced due to constructive interference, has been experimentally observed in a variety of systems~\cite{GrossHarochePhysRep1982}. For subradiant states the emission is suppressed owing to the destructive interference of the radiation from the emitters. Because of the inherently weak coupling of the subradiant states to external electromagnetic (EM) fields, their experimental studies have been limited. In the early experiments subradiant emission was observed for two trapped ions~\cite{DeVoe} as well as for two trapped molecules~\cite{Hettich}. Two-particle subradiant and superradiant states have an analogy with the gerade (even) and ungerade (odd) symmetry states of homonuclear molecular dimers, and subradiant states have also been created in weakly bound ultracold Sr$_2$~\cite{McGuyer} and Yb$_2$~\cite{Takasu} molecules. Superradiant states in dimers represent excitations via strong electric dipole transitions, while subradiant states may, e.g., be produced by weak magnetic dipole or electric quadrupole transitions.

Similar effects have been investigated in the context of plasmonics, where the analogy between nanostructured plasmonic resonators and molecular states encountered in natural media has lead to a plasmon hybridization theory~\cite{ProdanEtAlSCI2003}. Excitations in such systems, reminiscent of molecular wavefunctions, have consequently resulted in an analysis of dark and bright modes, with subradiant and superradiant characteristics, respectively. Narrow Fano resonances in the transmitted field or subradiant and superradiant excitations  were experimentally observed in plasmonic resonators consisting of three or four nanorods~\cite{LiuEtAlNatMat2009,Lovera}, and in plasmonic heptamers~\cite{FanCapasso,GiessenOligomers,Frimmer}, while efforts to increase the mode complexity of the resonators are attracting considerable attention~\cite{Dregely,Watson2016}. Recent theoretical work also highlighted that the connection between transmission resonances and the existence of subradiant excitations is less obvious than commonly recognized, since narrow Fano resonances are also produced by the interference of non-orthogonal modes even in the absence of subradiance~\cite{Hopkins13,Forestiere13}.

Experiments on EM field transmission in large planar metamaterial arrays demonstrated narrow spectral features and changes in the resonances due to the nature of
the resonators or the size of the system~\cite{papasimakis2009,FedotovEtAlPRL2010}. Such findings point toward a possible existence of subradiant excitations, and here we provide a detailed analysis of `coherent' planar metamaterial arrays that
 link the near- and far-field observations of the resonance behavior to large-scale numerical simulations of a microscopic theory of EM-field-mediated resonator interactions. We show that the observed resonance features in the reflection spectra
 directly correspond to the excitation of a single subradiant eigenmode spatially extending over the entire metamaterial lattice of over 1000 unit-cell resonators, or metamolecules. The results therefore rule out other possible explanations~\cite{Hopkins13,Forestiere13} of
 the narrow resonances as well as potential incoherent sources of suppressed radiation, such as radiation trapping~\cite{Holstein,Cummings86}, and also provide a post facto demonstration for the existence
 of subradiance in~\cite{papasimakis2009,FedotovEtAlPRL2010}.
 Rather surprisingly, we find that the created  \emph{multimetamolecule} subradiant state can confine 70\% (for the plasmonic case 60\%) of the total excitation of the array.
 Consequently, our analysis unambiguously demonstrates the existence of coherent and correlated many-body subradiant excitations that dramatically differ from subradiant modes restricted to a single individual metamolecule.
 The work not only provides a controlled environment for the study of many-body subradiance, but also a platform that can potentially be exploited, e.g., in high-precision measurements, metamaterial-based light-emitters~\cite{ZheludevEtAlNatPhot2008}, spectral filters~\cite{CAIT}, imaging~\cite{LemoultPRL10}, and nonlinear processes~\cite{Linden12}.

\begin{figure}
  \centering
  \includegraphics[width=1\columnwidth]{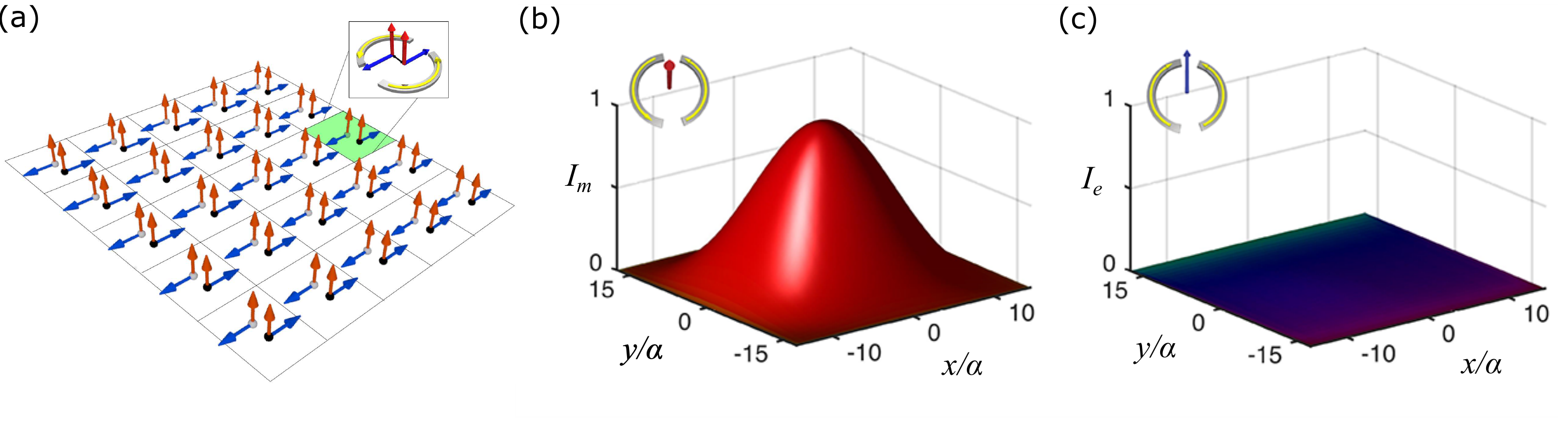}
  \caption{(a) A schematic illustration of the planar metamaterial array consisting of asymmetric split ring metamolecules. Each metamolecule has two constituent circuit resonator arcs, or meta-atoms, whose out-of-phase oscillating currents produce a strong net magnetic dipole perpendicular to the array. The arrows on the plane (blue arrows) represent the electric dipole of each arc and the arrows normal to the plane (red arrows) represent the magnetic dipoles generated by pairs of arcs. In a collective pure phase-coherent magnetic mode all magnetic dipoles in the array oscillate in-phase. (b-c) Numerically calculated magnetic dipole (b) and electric dipole (b) excitation profiles for this mode.
}   \label{fig:illustration}
 \end{figure}

We consider metamaterial planar arrays consisting of asymmetric split ring (ASR) metamolecules formed by two discrete circular arcs [or meta-atoms, see Fig.~\ref{fig:illustration}(a)] \cite{FedotovEtAlPRL2007,papasimakis2009,FedotovEtAlPRL2010}. Each ASR in the array, labeled by index $\ell$ ($\ell =1, \ldots, N$), can have a symmetric mode $(\ell,+)$, with the currents in the two arcs oscillating in-phase, and an antisymmetric mode $(\ell,-)$, with the currents oscillating $\pi$ out-of-phase. The symmetric mode produces a net electric dipole in the array plane and the antisymmetric mode a net magnetic dipole normal to the plane (accompanied by a weaker electric quadrupole moment).
We can extend the metamolecule analogy of molecular wavefunctions to the ASRs.
If the two arcs were symmetric, we would have a symmetric split-ring (SSR) resonator, and the metamolecule eigenfunctions would be similar to the wavefunctions of a  homonuclear dimer molecule. The quantum state of the dimer molecule reads
\beq
|\Psi_{\pm}\> = {1\over \sqrt{2}}\( |1,g;2,e\> \pm |1,e;2,g\>\)\,,
\eeq
where $g$ and $e$ denote the ground and excited states, respectively, and 1 and 2 are the two atoms of the dimer. Since the excited state has an odd symmetry, $|\Psi_{-}\> $ has an even symmetry and $|\Psi_{+}\> $ odd. Then the subradiant gerade $|\Psi_{-}\> $ corresponds to the antisymmetric ASR mode $(\ell,-)$ and the superradiant ungerade $|\Psi_{+}\> $ to the symmetric ASR mode $(\ell,+)$. The asymmetry in the lengths of the two ASR arcs shifts the resonance frequencies of the two meta-atoms and $(\ell,\pm)$ no longer represent the eigenmodes of the metamolecule. The ASR asymmetry couples the two modes $(\ell,\pm)$, such that both of them can in principle be excited by driving only one of them with incident EM fields (depending on the frequency, propagation direction, etc.). For an incident plane wave that propagates along the normal to the lattice couples directly only to the electric dipoles of the $(\ell,+)$ mode, since the magnetic dipoles point along the propagation direction. If the magnetic dipole radiation of the $(\ell,-)$ mode is much weaker than the electric dipole radiation of $(\ell,+)$, the asymmetry-induced coupling between broad and narrow resonance modes shows up as a characteristic Fano resonance in the transmission spectrum. However, in experimental situations the dipole radiation rates are comparable and no Fano resonance can be identified for a single ASR metamolecule~\cite{papasimakis2009}.
However, interactions between the resonators, mediated by scattered fields can have a profound impact.
In extreme cases the radiative interactions can lead to correlations between the excitations that are associated with recurrent scattering processes~\cite{Ishimaru1978,Morice1995a,Wiersma_recurrent,Ruostekoski1997a,Javanainen2014a,JavanainenMFT} in which a wave scatters more than once by the same resonator.

\begin{figure}
	\centering
	\includegraphics[width=1\columnwidth]{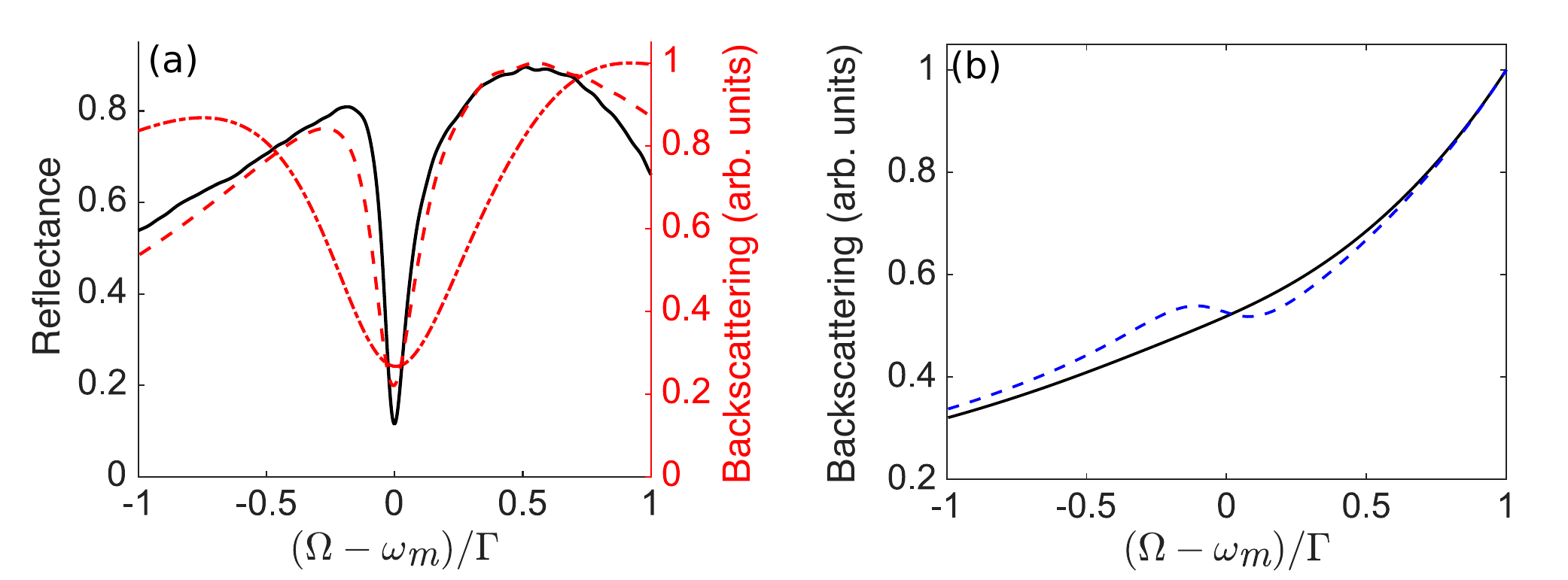}
	\caption{ (a) Experimentally measured reflectance (black line) and numerically calculated back-scattered intensity (red dashed line) spectra from a $30\times 36$ ASR microwave metamaterial array with a lattice spacing of $a=0.28\lambda$.  We also show calculated spectra (red dash-dot line) for arrays of plasmonic resonators ($a=0.2\lambda$). (b) Calculated spectra from microwave arrays of noninteracting metamolecules (black line), and arrays with a large lattice spacing of $a=1.9\lambda$ (dashed blue line) and thus weakened interactions. The frequencies are in the units of the single arc decay rate $\Gamma$, centered at the resonance frequency $\omega_m$ of the phase coherent magnetic eigenmode of the corresponding SSR array.}
	\label{fig:far_field_description}
\end{figure}

The collective response of ASR arrays is investigated by performing large-scale numerical simulations. We use the same general formalism as previously, with the details reported elsewhere~\cite{JenkinsLongPRB}, only a brief recap here~(Appendix). Each meta-atom $j$ is represented by a single mode of current oscillation that
behaves as an effective $RLC$ circuit with resonance frequency $\omega_j$. Each meta-atom is treated in the point dipole approximation with an in-plane electric dipole
$\spvec{d}_j(t)$  and a perpendicular
magnetic dipole $\spvec{m}_j(t)$; see Fig.~\ref{fig:illustration}.
The electric and magnetic dipole moments of the meta-atoms radiate at the rates $\Gamma_{\sje}$ and $\Gamma_{\sjm}$, respectively.
We also add a nonradiative loss rate $\Gamma_{\sjo}$, such that the total decay rate of the meta-atom excitations is $\Gamma=\Gamma_{\sje}+\Gamma_{\sjm}+\Gamma_{\sjo}$.
In the metamaterial array a meta-atom is driven by the sum of the incident fields  and the fields scattered by all the other meta-atom resonators in the system. The meta-atom then acts as a source of
radiation that, in turn, drives the other meta-atoms.
This leads to a coupled set of equations between the meta-atom excitations that describe the EM field mediated interactions and allow to evaluate the normal mode excitations of the system. This EM coupling between metamolecules leads to the emergence of many-body effects in the response of the metamaterial.
\begin{figure}
	\centering
	\includegraphics[width=1\columnwidth]{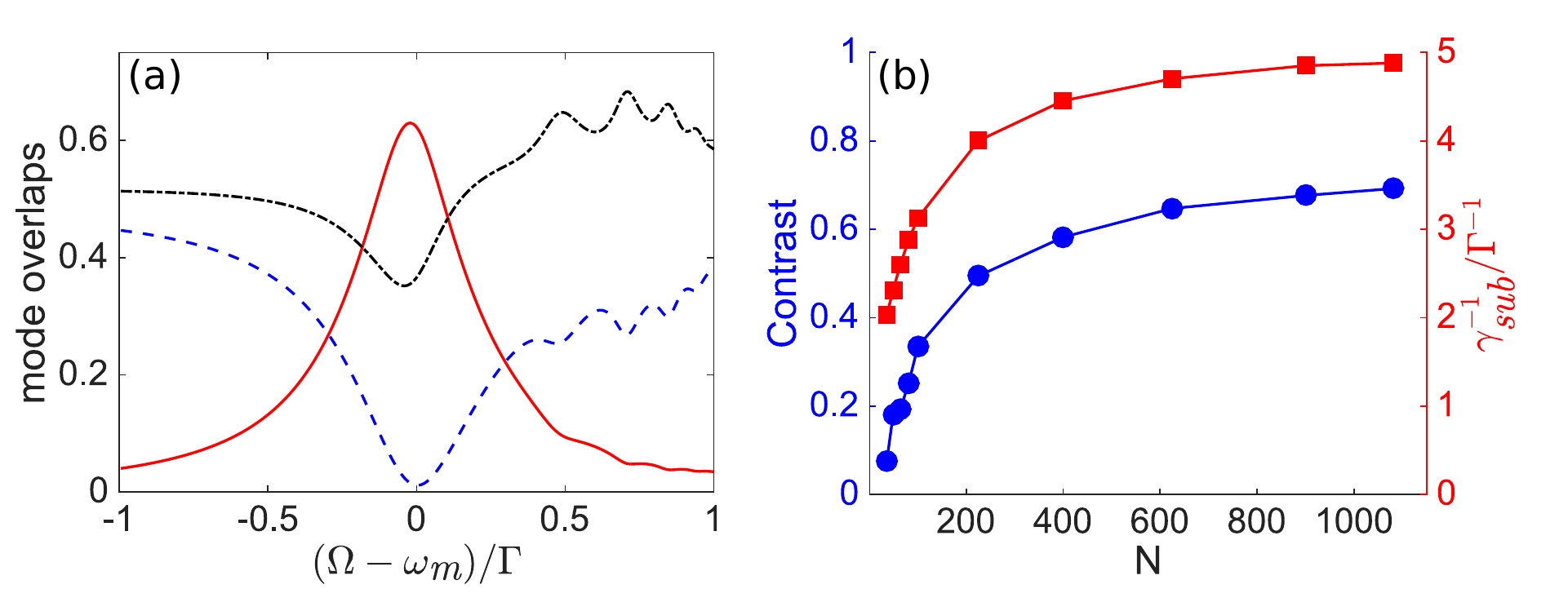}
	\caption{ (a) The contribution of a pure phase-coherent magnetic (solid red line) and a pure phase-coherent electric (dashed blue line) dipole mode to the steady-state excitation responsible for the spectrum in Fig.~2(a). We also show the proportion of all the other eigenmodes of the symmetric system (dash-dot black line). The pure phase-coherent magnetic dipole mode is dominant at the Fano resonance, while the pure phase-coherent electric dipole mode is significantly excited only outside the resonance. (b) The dependence of the Fano resonance depth (defined as $c = (I_{\rm max} -I_{\rm min})/I_{\rm max}$ where $I_{\rm min}$ is the minimum reflected intensity on resonance, and $I_{\rm max}$ is the reflected intensity at the lesser of the adjacent local maxima) (blue line) and inverse linewidth, $\gamma_{\rm sub}^{-1}$, of the dominant subradiant eigenmode of the ASR array (red line) on the size of the metamaterial array, $N$. In the absence of the Fano resonance $c=0$, while $c=1$ implies full reflection at the resonance frequency.
	}
	\label{fig:overlapscaling}
\end{figure}

Experimental setup is described in detail in~\cite{SavoEtAlPRB2012}. The measurements were performed on periodic metamaterial arrays of metallic ASR resonators. The asymmetry is introduced by a difference in length of the two arcs, corresponding to angles $160^\circ$ and $140^\circ$. This results in the different
resonance frequencies of the two arcs $\omega_0\pm\delta\omega$, where $\omega_0$ would be the resonance frequency of one arc
in an SSR metamolecule. The far-field characterization of the metamaterial arrays was performed in an anechoic chamber with broadband linearly polarized antennas at normal incidence. Near-field mapping of the metamaterial samples was performed in a microwave scanning-near field microscope \cite{SavoEtAlPRB2012}. Following the fabricated sample, the simulated microwave metamaterial array in a steady-state response comprised 30$\times$36 unit cells with a lattice spacing of $a=0.28\lambda\simeq 7.5$ mm ($\lambda=2\pi c/\omega_0$) assuming $\Gamma_{\sje}=\Gamma_{\sjm}$ and $\delta\omega=0.3\Gamma$. Any losses in the metamaterial are almost solely due to the supporting substrate, as metals at low frequencies (GHz) exhibit negligible dissipation loss. These were incorporate by setting $\Gamma_{\sjo}=0.07\Gamma$ that also provided the best fitting to the collective experimental response. In order to model the effects of the nonuniform illumination in the response of the array, we input the experimentally measured incident field profile in the numerical calculations.

In Fig.~2(a) we show a side-by-side comparison for the far-field measurements and numerical calculations of the reflected field intensity spectrum in a narrow cone in the back direction. The spectral response of the metamaterials exhibits a narrow Fano resonance \cite{FedotovEtAlPRL2007} associated with the magnetic dipole excitation of the metamolecules. Numerical calculations are in good qualitative agreement with the experimental observations, indicating that the model captures well the multiple scattering phenomena between the resonators. Although an isolated ASR metamolecule exhibits no sharp resonance, the large array of interacting metamolecules displays a high-quality \emph{collective} resonance. The resonance results entirely from interactions between the metamolecules that are mediated by the scattered fields. This is illustrated in Fig.~2(b), where we show the calculated spectra of weakly and noninteracting metamolecules of the same array, illustrating how an increased spatial separation leads to a substantially less pronounced, broader resonance.

The origin of these resonances can be traced to the eigenmodes of the metamaterial array. In particular, a uniform incident field normal to the lattice plane would couple most strongly to collective modes where metamolecules oscillate in phase, which is the case for a pure phase-coherent electric (PE) and a pure phase-coherent magnetic (PM) dipole mode. For a SSR array these are collective eigenmodes of the system, similarly as the $(\ell,\pm)$ modes are eigenstates of a single SSR metamolecule. The PM mode of the studied case is shown in Fig.~1(b). Owing to the asymmetry of the ASR arcs, PE and PM modes in the ASR metamaterial array are no longer eigenmodes and are coupled by the asymmetry.
The role of the different modes can be quantified by analyzing the collective eigenmodes of the strongly coupled resonator array. In Fig.~3(a) we show the overlap between the PE and PM modes and the steady-state excitation responsible for the far-field spectrum of Fig.~2(a). Here the overlap measure between an eigenmode $\colvec{v}_j$ with an excitation $\colvec{b}$ is defined by
$
  O_{j}(\colvec{b}) \equiv  | \colvec{v}_j^T \colvec{b} |^2
  /\sum_i | \colvec{v}_i^T \colvec{b} |^2
$,
where the summation runs over all the eigenmodes. Since the incident field in the experiment is not uniform, the coupling can drive strongly also other modes than PE and PM modes.
However, the numerical results indicate that both PM and PE modes still play a significant role in the response of the metamaterial.
PM mode excitation constitutes 63\% of the total excitation at the resonance and rapidly decays outside of it. PE excitation is notable only outside of the resonance. The most remarkable feature is the very strongly subradiant nature of PM mode; we find that in the corresponding SSR array, where the symmetry between the arcs of the metamolecules is not broken and where PM mode is an eigenmode, its radiative decay rate would only be about $\gamma_m\simeq 0.011\Gamma$ (together with the nonradiative ohmic loss rate, the total decay rate still only 0.081$\Gamma$). For PE mode the total decay rate in the corresponding SSR array would be about $\gamma_e\simeq 3.0\Gamma$, indicating superradiant decay. In the ASR array, the asymmetry between the ASR arcs couples PE and PM modes. Hence, the Fano resonance at the frequency $\omega_m$ (the resonance frequency of PM mode) results from the interference between the collective subradiant PM mode with an extremely narrow radiative linewidth and the superradiant PE mode. The general behavior of PM and PE modes is consistent with their radiation patterns. The dipoles aligned in the plane in PE mode strongly reflect EM fields normal to the plane, while PM mode dipoles emit into the plane of the lattice and suppress reflection.

So far we have described the ASR metamaterial response in terms of PM and PE modes that are not eigenmodes in the ASR array. In order to show that we have prepared subradiant many-body excitations we need to calculate the eigenmodes of the ASR array~(Appendix). The decay of a radiative excitation amplitude then satisfies $\sum_j b_j \exp (-\gamma_j t)$, where $\gamma_j$ are the collective eigenmode linewidths and $|b_j|^2$ are the occupation estimates.
We find that the steady-state excitation at the Fano resonance is overwhelmingly dominated (close to 70\% of the total excitation) by a subradiant eigenmode with the decay rate of  $\gamma_{\rm sub}\simeq 0.21\Gamma$ and the resonance frequency $\omega_{\rm sub}\simeq\omega_m-0.017\Gamma$. Remarkably, this subradiant excitation is a correlated many-body excitation between a large number of metamolecules and extends over the entire metamaterial lattice. This is illustrated in Fig.~\ref{fig:overlapscaling}(b), where we show the numerically calculated dependence of the radiative linewidth of the eigenmode on the size of the array. In Fig.~\ref{fig:overlapscaling}(b) we approximately maintain the aspect ratio of the array while changing the number of metamolecules from one to the experimental value of 1080.
The increase in the number of resonators notably continues reducing the linewidth even in the case of over 1000 metamolecules (over 2000 meta-atoms). Figure~3(b) also shows how the far-field resonance properties are directly linked to the radiative resonance linewidth of the subradiant excitation by comparing the resonance contrast with the eigenmode linewidth; we observe notably similar profiles for the emergent resonance and the subradiant mode linewidth as a function of the number of metamolecules, indicating that both result from the same \emph{collective} interaction phenomena. The transmission resonance through an ASR array and its narrowing as a function of the size of the system has been previously experimentally observed~\cite{FedotovEtAlPRL2010}.
The emergence of the Fano resonance implies a coupling between modes with a broad and a narrow resonance. Although this does not necessarily indicate the existence of subradiance in the system, our detailed theoretical and numerical comparisons provide strong evidence of correlated many-body multimetamolecule subradiant excitations of distant metamolecules that spatially extend over the entire metamaterial lattice.

\begin{figure}
	\centering
	\includegraphics[width=1\columnwidth]{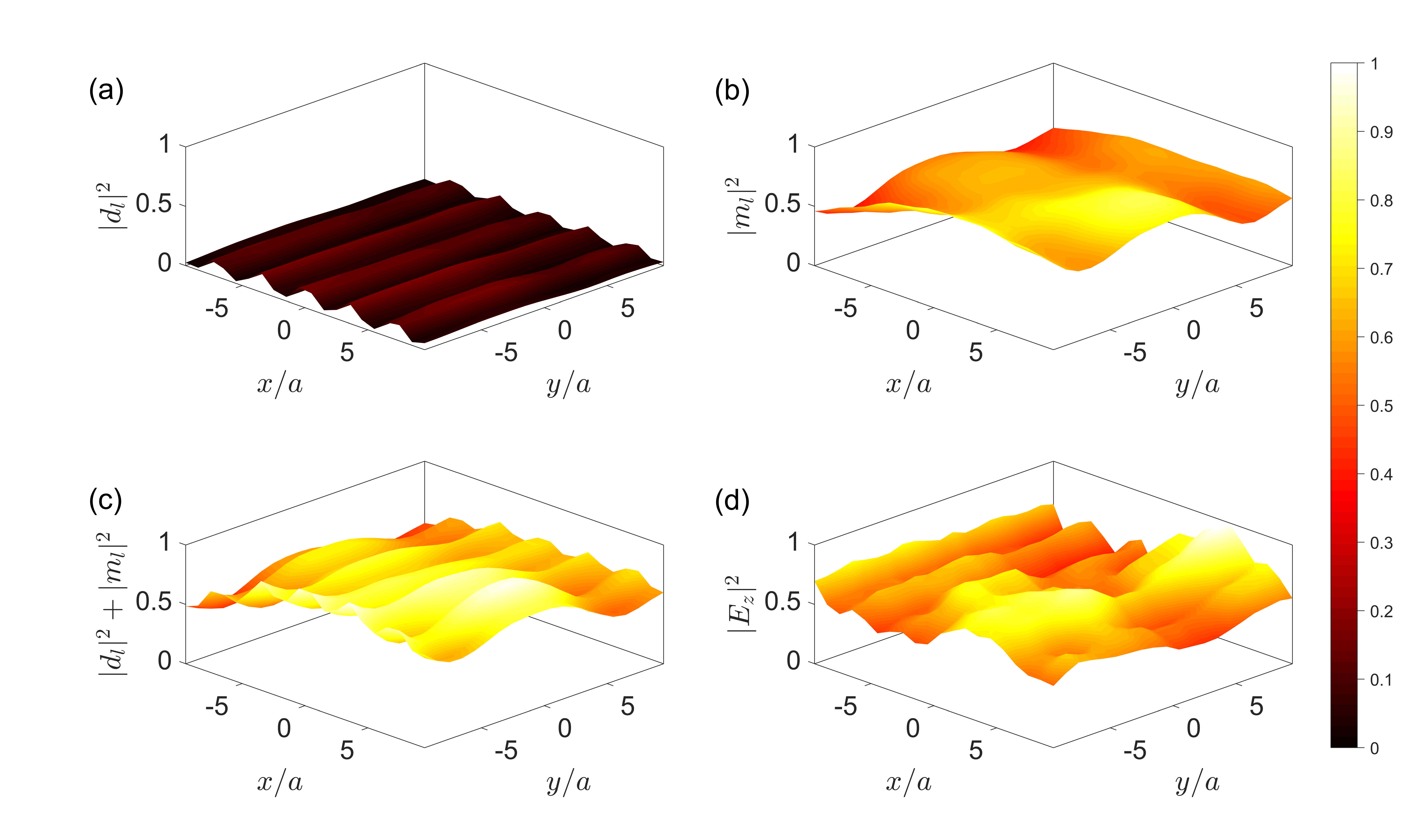}
	\caption{ Numerically calculated (a-c) and experimental (d) near-field excitations of a microwave ASR metamaterial array at the transmission peak: (a) electric dipole intensity   $|d_\ell|^2$; (b)  magnetic dipole intensity $|m_\ell|^2$; (c) the total excitation energy $|d_\ell|^2 + |m_\ell|^2$; and (d) experimentally measured electric field intensity.
	}
	\label{fig:near_field_ordered_exp_th_comp}
\end{figure}

In Fig.~\ref{fig:near_field_ordered_exp_th_comp} we show near-field measurements of the microwave radiation of the array at the Fano resonance and the corresponding theoretical calculation.
Using the experimentally measured nonuniform incident field profile, the numerical model qualitatively captures the characteristic stripelike feature of the near-field excitation along the axis of the ASR arcs, but underestimates the nonuniformity of the excitations. In the theoretical model we also analyze the separate contributions of the magnetic and electric dipole excitations.
The stripelike pattern is identifiable only in the electric dipole excitations, but also the near field displays the concentration of resonant excitation on the magnetic dipoles and PM mode.

One may ask whether a similar multimetamolecule subradiant excitation can be observed also in plasmonic metamaterials in the optical domain, where ohmic losses in metals are higher than at microwave frequencies. In plasmonic resonators the stronger ohmic losses result in absorption of light and suppress the long-range light-mediated interactions between the different metamolecules. By performing numerical simulations for a plasmonic ASR array in the optical domain using realistic parameters we found that suitable parameter regimes for strong collective effects can also be found for plasmonic systems when the radiative decay is sufficiently strong ($Q$-factors of individual resonators are sufficiently low).
One can show that a cooperative resonance is especially pronounced if the asymmetry that drives the subradiant mode also satisfies $\delta\omega^2 \gg \gamma_m\gamma_e$, requiring
large $\delta\omega$ when the nonradiative losses are substantial.
For instance, we take $\Gamma_{\sjo}=0.25\Gamma$ that is comparable with those observed for Fano resonance experiments on gold rods~\cite{LiuEtAlNatMat2009} and obtained by Drude-model based estimates~\cite{kuwate}. The results for asymmetry $\delta\omega=0.75\Gamma$ and lattice spacing $a=0.2\lambda$ are displayed in Fig.~2(a) that clearly show the existence of the resonance in the far-field spectrum. The resonance is broader than in the microwave case, but still includes a strong contribution from the PM mode ($\sim 45\%$), whereas the PE mode is at a minimum ($2.5\%$)~(Appendix). We also calculated the ASR eigenmodes, and at the exact resonance $\sim60\%$ of the excitation is confined in a subradiant eigenmode with the linewidth of  $\gamma_{\rm sub}\simeq0.75\Gamma$, indicating a dominant collective subradiant excitation in the system. (In the corresponding SSR system the resonance linewidths of PM and PE modes would be $\gamma_m\simeq0.28\Gamma$ and $\gamma_e\simeq 4.7\Gamma$.)

In conclusion, we showed that a planar metamaterial array can be designed in such a way that the excitation energy is overwhelmingly dominated by a subradiant eigenmode that spatially extends over the large array.
This is very different, e.g., from recent observations of subradiance~\cite{Guerin_subr16} in an atomic vapor where only a very small fraction of the emitters was found to possess a suppressed decay rate. Our analysis of the controlled state preparation paves the way
towards engineering complex correlated EM excitations that consists of large numbers of resonators, with potential applications, e.g., in light storage, optical memories, and light emission.
The metamaterial resonator arrays also bear resemblance to other resonant emitter lattices, such as cold-atom systems~\cite{Jenkins2012a,Bettles2014} which similarly respond to light as classical oscillators in the typically applied low light intensity limit~\cite{Ruostekoski1997a,Lee16}. However, finding experimental evidence of correlated light-mediated interactions in atomic vapors is generally challenging~\cite{Jennewein_trans}, and correlated light excitations could therefore potentially be better utilized in metamaterial applications.

\begin{acknowledgments}
We acknowledge financial support from the EPSRC (EP/G060363/1, EP/M008797/1), the Leverhulme Trust, the Royal Society, and the MOE Singapore
Grant No. MOE2011-T3-1-005. We also acknowledge the use of the IRIDIS
High Performance Computing Facility at the University of
Southampton.
\end{acknowledgments}

\appendix

\setcounter{equation}{0}
\setcounter{figure}{0}
\renewcommand{\theequation}{A\arabic{equation}}
\renewcommand{\thefigure}{A\arabic{figure}}

\section{Appendices}

\subsection*{Numerical model}

\subsubsection*{Radiative coupling between the resonators}

We briefly describe the numerical simulations of the electromagnetic (EM) response of a planar metamaterial array.
The general formalism to describe interacting magnetodielectric resonators is presented in \cite{JenkinsLongPRB}. For the rectangular array of asymmetric split ring (ASR) metamolecules we represent each meta-atom $j$ by a single mode of current oscillation that
behaves as an effective $RLC$ circuit with resonance frequency $\omega_j$. We label the meta-atoms by indices $j$ ($j = 1,\ldots,2N$) such that the $\ell$th ASR metamolecule includes the two meta-atoms $2\ell -1$ and $2\ell$.
In a symmetric split ring (SSR) metamolecule the resonance frequencies of the two meta-atoms are equal $\omega_0$, but in the ASRs the symmetry is broken by adjusting the lengths of the resonators in such a way that the resonance frequencies of the two arcs in each ASR metamolecule become $\omega_0\pm \delta\omega$.

In the following discussion for the radiative interactions between the resonators, all the field and resonator amplitudes refer to the slowly-varying versions of the
positive frequency components of the corresponding variables,
where the rapid oscillations $e^{-i\Omega t}$ ($k=\Omega/c$) due to the frequency, $\Omega$, of the incident wave have been factored out in the rotating wave approximation.
The current excitations in the meta-atoms interact with the propagating field and radiate electric and magnetic fields. Each meta-atom is treated in the point dipole approximation with an electric dipole
$\spvec{d}_j(t) = d_j(t)\unitvec{e}_y$  and
magnetic dipole $\spvec{m}_j(t) = m_j(t)\unitvec{m}_j$, where $\unitvec{m}_{2\ell} = -\unitvec{m}_{2\ell-1} \equiv \unitvec{m} = \unitvec{e}_z$; see Fig.~1 in the main section.
The normal mode amplitude of the current excitations in the meta-atom $j$ then reads~\cite{JenkinsLongPRB}
\begin{equation}
  \label{eq:b_def}
  b_j(t) = \(\frac{k^3}{12\pi\epsilon_0}\)^{1/2}\left(
    \frac{d_j(t)}{\sqrt{\Gamma_{\sje}}} + i
    \frac{m_j(t)}{c\sqrt{\Gamma_{\sjm}}}\right) \,.
\end{equation}
Here $\Gamma_{\sje}$ and $\Gamma_{\sjm}$ denote the electric and magnetic dipole decay rates of an isolated meta-atom, respectively. The total decay rate of the meta-atom excitations is then $\Gamma=\Gamma_{\sje}+\Gamma_{\sjm}+\Gamma_{\sjo}$, where we have also added a nonradiative loss rate $\Gamma_{\sjo}$.

Moreover, the scattered fields are given by ${\bf E}_{S}=\sum_j {\bf E}_{S}^{(j)}$ and ${\bf H}_{S}=\sum_j {\bf H}_{S}^{(j)}$ where the contributions from the meta-atom $j$ read
\commentout{
\begin{align}
{\bf E}_{\text{sc},j}({\bf r},t)&=\frac{k^3}{4\pi\epsilon_0}\sum_j
\bigg[{\sf G}({\bf r} -{\bf r}_j){\bf d}_j(t)\nonumber\\
&+\frac{1}{c}{\sf G}_\times({\bf r} - {\bf r}_j)
{\bf m}_j(t)\bigg],\label{eq:Esc}\\
{\bf B}_{\text{sc},j}({\bf r},t)&=\frac{\mu_0k^3}{4\pi}\sum_j
\bigg[{\sf G}({\bf r} -{\bf r}_j){\bf m}_j(t)\nonumber\\
&-c{\sf G}_\times({\bf r} - {\bf r}_j){\bf d}_j(t)\bigg],
\label{eq:Bsc}
\end{align}
}
\begin{align}
{\bf E}_{S}^{(j)}({\bf r},t)&=\frac{k^3}{4\pi\epsilon_0}
\bigg[{\sf G}({\bf r} -{\bf r}_j){\bf d}_j
+\frac{1}{c}{\sf G}_\times({\bf r} - {\bf r}_j)
{\bf m}_j\bigg],\label{eq:Esc}\\
{\bf H}_{S}^{(j)}({\bf r},t)&=\frac{k^3}{4\pi}
\bigg[{\sf G}({\bf r} -{\bf r}_j){\bf m}_j
-c{\sf G}_\times({\bf r} - {\bf r}_j){\bf d}_j\bigg]\,.
\label{eq:Bsc}
\end{align}
The radiation kernel ${\sf G}({\bf r} - {\bf r}_j)$ determines
the electric (magnetic) field at ${\bf r}$, from an oscillating electric (magnetic) dipole of the meta-atom $j$
at ${\bf r}_j$~\cite{Jackson}. The cross kernel ${\bf G}_\times({\bf r} - {\bf r}_j)$ describes
 the electric (magnetic) field at ${\bf r}$ an oscillating magnetic (electric) dipole
at ${\bf r}_j$.

Each meta-atom is driven by the incident fields, $\cbE_0(\spvec{r},t)$ and $\cbH_0(\spvec{r},t)$,  and the fields scattered by all the other resonators in the system,
\begin{align}
{\bf E}_{\text{ext}}({\bf r}_j,t)& = \cbE_0(\spvec{r},t)
+ \sum_{l\ne j}{\bf E}_{S}^{(l)}({\bf r},t),
\label{eq:Eext}\\
{\bf H}_{\text{ext}}({\bf r}_j,t)& = \cbH_0(\spvec{r},t) +
\sum_{l\ne j}{\bf H}_{S}^{(l)}({\bf r},t)\,,
\label{eq:Bext}
\end{align}
where the scattered fields are given by Eqs.~\eqref{eq:Esc} and~\eqref{eq:Bsc}.  The EM fields couple to the current excitations of the meta-atom via its electric and magnetic dipole moments according to Eq.~\eqref{eq:b_def}.

Collecting these together, we obtain a coupled set of equations between the meta-atom excitations that describe the EM field mediated interactions. In terms of
$\colvec{b} \equiv (b_1, b_2, \ldots, b_{2N})^T$,
we may write them as~\cite{JenkinsLongPRB}
\beq
\dot{\colvec{b}} = \mathcal{C}\colvec{b} + \colvec{F}(t)\,.
\label{eqnmotion}
\eeq
The off-diagonal elements of the matrix $\mathcal{C}$ describe radiative interactions between the different arcs mediated by the scattered field that incorporate the retardation effects with short- and long-range interactions. The diagonal  elements represent the damped free oscillations of the arcs. The driving of the meta-atom by the incident field is encapsulated  in $\colvec{F}(t)$. Evaluating the normal mode excitations of the system and the corresponding scattered fields~\eqref{eq:Esc} and~\eqref{eq:Bsc} then yield the metamaterial array's response.

\subsubsection*{Steady-state response}

In the numerical simulations of the planar metamaterial illuminated by the driving fields, we solve the steady-state response,
\beq
\colvec{b}=-\mathcal{C}^{-1} \colvec{F}\,,
\eeq
obtained from \eqref{eqnmotion}.

In our simulations we consider the experimental arrangement of the 30$\times$36  array of ASR metamolecules, with the single metamolecule magnetic and electric decay rates satisfying $\Gamma_{\sje}=\Gamma_{\sjm}$.
In the microwave ASR system the nonradiative losses are almost solely due to the supporting substrate and we incorporate these in the numerical model by setting $\Gamma_{\sjo}=0.07\Gamma$.
For the plasmonic system the nonradiative loss rate $\Gamma_{\sjo}=0.25\Gamma$ is chosen to be comparable with those observed for Fano resonance experiments on gold rods~\cite{LiuEtAlNatMat2009}
and obtained by Drude-model based estimates~\cite{kuwate}.

For the microwave (plasmonic) case we consider a lattice spacing $a=0.28\lambda$ ($a=0.2\lambda$),  the separation of the two meta-atoms in each metamolecule $0.114\lambda$ ($0.075\lambda$),
and the asymmetry between the two arcs of a metamolecule
$\delta\omega=0.3\Gamma$  ($\delta\omega=0.7\Gamma$).

In order to model the effects of the nonuniform illumination in the response of the array, we input the experimentally measured incident field profile in the numerical calculations.

\subsubsection*{Eigenmode calculations}

In a metamaterial array, we have a system of $N$ ASR meta-molecules, or $2N$ single-mode resonators (meta-atoms). These possess $2N$ collective modes of current oscillation, with corresponding collective resonance frequencies and decay rates.
We calculate the eigenmodes of the entire 30$\times$36 planar metamaterial system of  2160 interacting meta-atoms by diagonalizing the matrix $\mathcal{C}$. In Fig.~3(b) of the main section we also calculate the eigenmodes by varying the size of the array.
We generally consider the two cases with the array consisting of (i) SSR metamolecules and (ii) ASR metamolecules. The description of the response in terms of the eigenmodes of the SSR array (i)  is useful in explaining the emergent Fano resonance when the system becomes more strongly interacting. Specifically, the Fano resonance results from the destructive interference between the pure phase-coherent electric (PE) and pure phase-coherent magnetic (PM) dipole modes. Diagonalizing the system (ii), on the other hand, provides the true eigenmodes of the system and shows how the steady-sate of excitation of the array is dominated by the spatially extended multimetamolecule subradiant mode.

In the main text the emphasis is on the superradiant and subradiant modes that participate most strongly in the response. However, among the rest of the modes there are
also eigenmodes with notably narrower linewidths. For example, for the SSR system the eigenmodes exhibit a broad distribution of resonance frequencies and decay rates. We find that the largest decay rate in the system, corresponding to the most superradiant mode, is about 14.5$\Gamma$. The smallest decay rate is negligibly small compared with nonradiative losses with the value of $9.4\times 10^{-10}\Gamma$. Due to extremely weak coupling of this mode to external fields, its excitation is very challenging.

\subsection*{Eigenmode contributions of the excitations }
\begin{figure}
  \centering
  \includegraphics[width=1\columnwidth]{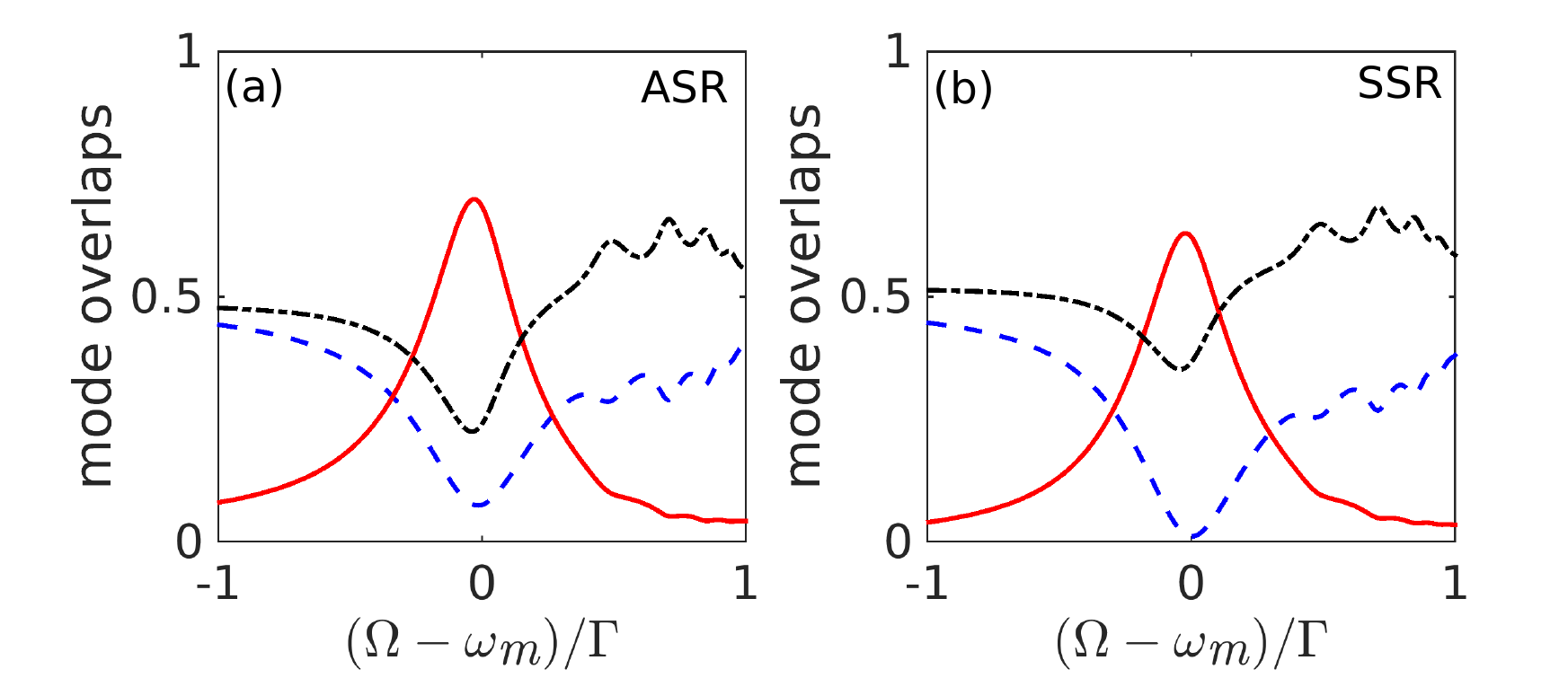}
  \caption{Contributions of dominantly electric (solid red line) and magnetic (dashed blue line) eigenmodes to the steady-state excitation of microwave ASR arrays under a decomposition in ASR (a) and SSR (b) eigenmodes. The sum of the contributions of all the other modes is marked by a dashed-dotted black line in both panels.
  }
  \label{fig:muwave}
\end{figure}
\begin{figure}
  \centering
  \includegraphics[width=1\columnwidth]{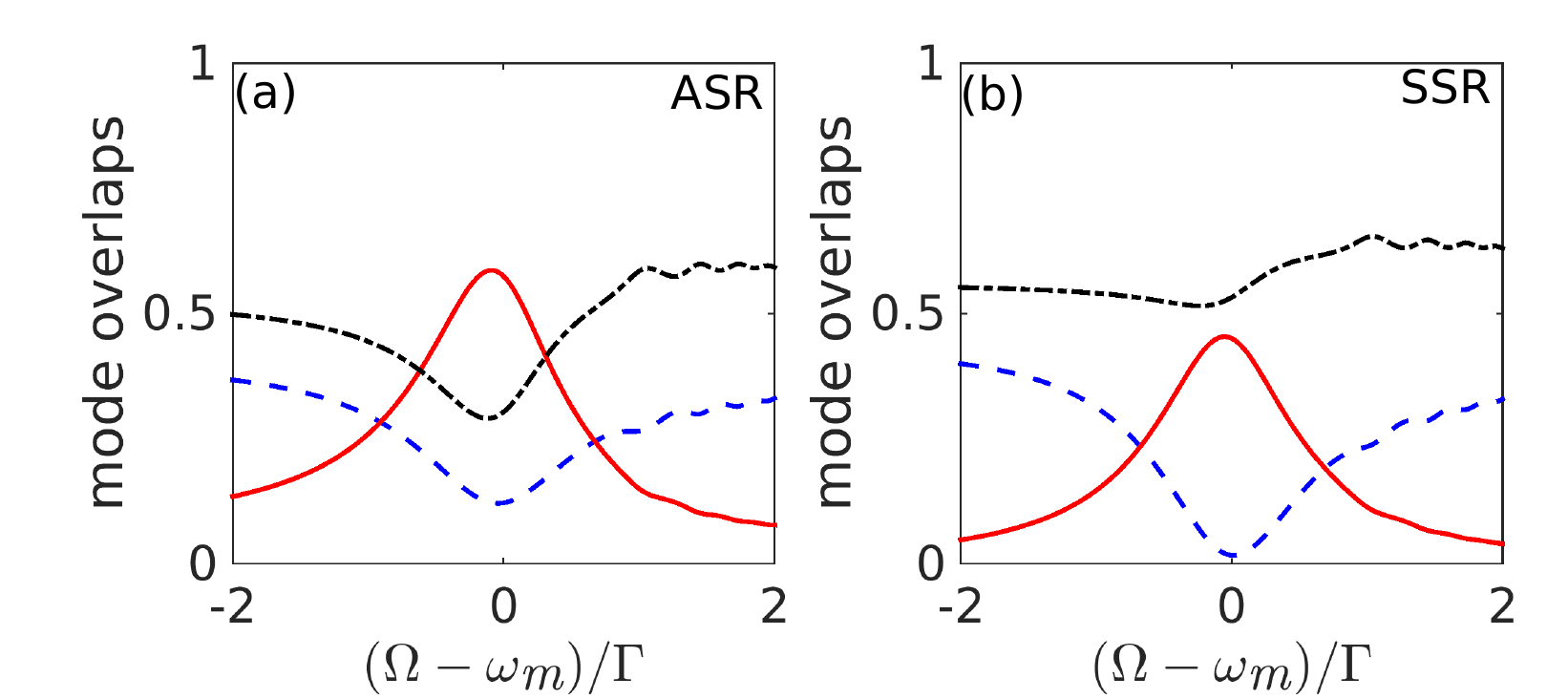}
  \caption{Contributions of dominantly electric (solid red line) and magnetic (dashed blue line) eigenmodes to the steady-state excitation of plasmonic ASR arrays under a decomposition in ASR (a) and SSR (b) eigenmodes. The sum of the contributions of all the other modes is marked by a dashed-dotted black line in both panels.
  }
  \label{fig:plasmonic}
\end{figure}
The different eigenmode contributions of the steady-state excitations were shown in the main section of the paper, where PM and PE modes were shown to characterize the Fano transmission resonance. In Fig.~\ref{fig:muwave} we compare the occupations of the steady-state excitation in the two eigenmode basis, one formed by the interacting SSR metamolecules and the other one by the interacting ASR metamolecules. The SSR one is the same as Fig.~3(a) in the main section. The ASR curves illustrate the overwhelmingly dominant excitation of the subradiant eigenmode of the ASR array. In Fig.~\ref{fig:plasmonic}, we present the corresponding occupations for plasmonic metamaterial arrays. Similarly to the microwave case, the subradiant eigenmode dominates over the coherent electric dipole mode, although other modes provide also provide strong contributions.

\subsection*{Experimental setup and samples}
Asymmetrically-split ring (ASR) metamaterial arrays were manufactured  by etching a $35$ $\mu$m copper cladding on an FR4 printed circuit board (PCB) substrate. The thickness of the substrate is $1.6$ mm, whereas its permittivity is $\epsilon\simeq 4.5+0.15i$.  Each ASR has an inner and outer radius of $2.8$ and $3.2$ mm, respectively, with $\Gamma=3.5~\mathrm{GHz}$. The asymmetry is introduced by a difference in length of the two arcs, corresponding to angles $160^\circ$ and $140^\circ$. The array consists of $30\times 36$ unit cells, each with dimensions of $7.5\times 7,5$ mm$^2$.

Transmission and reflectivity measurements were performed in an anechoic chamber with broadband linearly polarized antennas  (Schwarzbeck BBHA 9120D) at normal incidence, where the electric field amplitude and phase was recorded by a vector network analyzer  (Agilent E8364B)  in the range of $2-18$ GHz. The polarization of the incident wave was normal to the plane of symmetry of the ASR. Near-field mapping of the metamaterial sample was performed in a microwave scanning-near field microscope,  where an electric monopole  with a length of $2.5$ mm was mounted on a motorized stage and collected the near electric field at a distance of $1$ mm from the metamaterial. The spatial resolution in the sample plane was $1$ mm.


\begin{thebibliography}{40}%
\makeatletter
\providecommand \@ifxundefined [1]{%
 \@ifx{#1\undefined}
}%
\providecommand \@ifnum [1]{%
 \ifnum #1\expandafter \@firstoftwo
 \else \expandafter \@secondoftwo
 \fi
}%
\providecommand \@ifx [1]{%
 \ifx #1\expandafter \@firstoftwo
 \else \expandafter \@secondoftwo
 \fi
}%
\providecommand \natexlab [1]{#1}%
\providecommand \enquote  [1]{``#1''}%
\providecommand \bibnamefont  [1]{#1}%
\providecommand \bibfnamefont [1]{#1}%
\providecommand \citenamefont [1]{#1}%
\providecommand \href@noop [0]{\@secondoftwo}%
\providecommand \href [0]{\begingroup \@sanitize@url \@href}%
\providecommand \@href[1]{\@@startlink{#1}\@@href}%
\providecommand \@@href[1]{\endgroup#1\@@endlink}%
\providecommand \@sanitize@url [0]{\catcode `\\12\catcode `\$12\catcode
  `\&12\catcode `\#12\catcode `\^12\catcode `\_12\catcode `\%12\relax}%
\providecommand \@@startlink[1]{}%
\providecommand \@@endlink[0]{}%
\providecommand \url  [0]{\begingroup\@sanitize@url \@url }%
\providecommand \@url [1]{\endgroup\@href {#1}{\urlprefix }}%
\providecommand \urlprefix  [0]{URL }%
\providecommand \Eprint [0]{\href }%
\providecommand \doibase [0]{http://dx.doi.org/}%
\providecommand \selectlanguage [0]{\@gobble}%
\providecommand \bibinfo  [0]{\@secondoftwo}%
\providecommand \bibfield  [0]{\@secondoftwo}%
\providecommand \translation [1]{[#1]}%
\providecommand \BibitemOpen [0]{}%
\providecommand \bibitemStop [0]{}%
\providecommand \bibitemNoStop [0]{.\EOS\space}%
\providecommand \EOS [0]{\spacefactor3000\relax}%
\providecommand \BibitemShut  [1]{\csname bibitem#1\endcsname}%
\let\auto@bib@innerbib\@empty
\bibitem [{\citenamefont {Dicke}(1954)}]{Dicke1954}%
  \BibitemOpen
  \bibfield  {author} {\bibinfo {author} {\bibfnamefont {R.~H.}\ \bibnamefont
  {Dicke}},\ }\href@noop {} {\bibfield  {journal} {\bibinfo  {journal} {Phys.
  Rev.}\ }\textbf {\bibinfo {volume} {93}},\ \bibinfo {pages} {99} (\bibinfo
  {year} {1954})}\BibitemShut {NoStop}%
\bibitem [{\citenamefont {Gross}\ and\ \citenamefont
  {Haroche}(1982)}]{GrossHarochePhysRep1982}%
  \BibitemOpen
  \bibfield  {author} {\bibinfo {author} {\bibfnamefont {M.}~\bibnamefont
  {Gross}}\ and\ \bibinfo {author} {\bibfnamefont {S.}~\bibnamefont
  {Haroche}},\ }\href@noop {} {\bibfield  {journal} {\bibinfo  {journal} {Phys.
  Rep.}\ }\textbf {\bibinfo {volume} {93}},\ \bibinfo {pages} {301} (\bibinfo
  {year} {1982})}\BibitemShut {NoStop}%
\bibitem [{\citenamefont {DeVoe}\ and\ \citenamefont {Brewer}(1996)}]{DeVoe}%
  \BibitemOpen
  \bibfield  {author} {\bibinfo {author} {\bibfnamefont {R.~G.}\ \bibnamefont
  {DeVoe}}\ and\ \bibinfo {author} {\bibfnamefont {R.~G.}\ \bibnamefont
  {Brewer}},\ }\href {\doibase 10.1103/PhysRevLett.76.2049} {\bibfield
  {journal} {\bibinfo  {journal} {Phys. Rev. Lett.}\ }\textbf {\bibinfo
  {volume} {76}},\ \bibinfo {pages} {2049} (\bibinfo {year}
  {1996})}\BibitemShut {NoStop}%
\bibitem [{\citenamefont {Hettich}\ \emph {et~al.}(2002)\citenamefont
  {Hettich}, \citenamefont {Schmitt}, \citenamefont {Zitzmann}, \citenamefont
  {Kühn}, \citenamefont {Gerhardt},\ and\ \citenamefont
  {Sandoghdar}}]{Hettich}%
  \BibitemOpen
  \bibfield  {author} {\bibinfo {author} {\bibfnamefont {C.}~\bibnamefont
  {Hettich}}, \bibinfo {author} {\bibfnamefont {C.}~\bibnamefont {Schmitt}},
  \bibinfo {author} {\bibfnamefont {J.}~\bibnamefont {Zitzmann}}, \bibinfo
  {author} {\bibfnamefont {S.}~\bibnamefont {Kühn}}, \bibinfo {author}
  {\bibfnamefont {I.}~\bibnamefont {Gerhardt}}, \ and\ \bibinfo {author}
  {\bibfnamefont {V.}~\bibnamefont {Sandoghdar}},\ }\href {\doibase
  10.1126/science.1075606} {\bibfield  {journal} {\bibinfo  {journal}
  {Science}\ }\textbf {\bibinfo {volume} {298}},\ \bibinfo {pages} {385}
  (\bibinfo {year} {2002})}\BibitemShut {NoStop}%
\bibitem [{\citenamefont {McGuyer}\ \emph {et~al.}(2015)\citenamefont
  {McGuyer}, \citenamefont {McDonald}, \citenamefont {Iwata}, \citenamefont
  {Tarallo}, \citenamefont {Skomorowski}, \citenamefont {Moszynski},\ and\
  \citenamefont {Zelevinsky}}]{McGuyer}%
  \BibitemOpen
  \bibfield  {author} {\bibinfo {author} {\bibfnamefont {B.~H.}\ \bibnamefont
  {McGuyer}}, \bibinfo {author} {\bibfnamefont {M.}~\bibnamefont {McDonald}},
  \bibinfo {author} {\bibfnamefont {G.~Z.}\ \bibnamefont {Iwata}}, \bibinfo
  {author} {\bibfnamefont {M.~G.}\ \bibnamefont {Tarallo}}, \bibinfo {author}
  {\bibfnamefont {W.}~\bibnamefont {Skomorowski}}, \bibinfo {author}
  {\bibfnamefont {R.}~\bibnamefont {Moszynski}}, \ and\ \bibinfo {author}
  {\bibfnamefont {T.}~\bibnamefont {Zelevinsky}},\ }\href {\doibase
  10.1038/NPHYS3182} {\bibfield  {journal} {\bibinfo  {journal} {Nat. Phys.}\
  }\textbf {\bibinfo {volume} {11}},\ \bibinfo {pages} {32} (\bibinfo {year}
  {2015})}\BibitemShut {NoStop}%
\bibitem [{\citenamefont {Takasu}\ \emph {et~al.}(2012)\citenamefont {Takasu},
  \citenamefont {Saito}, \citenamefont {Takahashi}, \citenamefont {Borkowski},
  \citenamefont {Ciury\l{}o},\ and\ \citenamefont {Julienne}}]{Takasu}%
  \BibitemOpen
  \bibfield  {author} {\bibinfo {author} {\bibfnamefont {Y.}~\bibnamefont
  {Takasu}}, \bibinfo {author} {\bibfnamefont {Y.}~\bibnamefont {Saito}},
  \bibinfo {author} {\bibfnamefont {Y.}~\bibnamefont {Takahashi}}, \bibinfo
  {author} {\bibfnamefont {M.}~\bibnamefont {Borkowski}}, \bibinfo {author}
  {\bibfnamefont {R.}~\bibnamefont {Ciury\l{}o}}, \ and\ \bibinfo {author}
  {\bibfnamefont {P.~S.}\ \bibnamefont {Julienne}},\ }\href {\doibase
  10.1103/PhysRevLett.108.173002} {\bibfield  {journal} {\bibinfo  {journal}
  {Phys. Rev. Lett.}\ }\textbf {\bibinfo {volume} {108}},\ \bibinfo {pages}
  {173002} (\bibinfo {year} {2012})}\BibitemShut {NoStop}%
\bibitem [{\citenamefont {Prodan}\ \emph {et~al.}(2003)\citenamefont {Prodan},
  \citenamefont {Radloff}, \citenamefont {Halas},\ and\ \citenamefont
  {Nordlander}}]{ProdanEtAlSCI2003}%
  \BibitemOpen
  \bibfield  {author} {\bibinfo {author} {\bibfnamefont {E.}~\bibnamefont
  {Prodan}}, \bibinfo {author} {\bibfnamefont {C.}~\bibnamefont {Radloff}},
  \bibinfo {author} {\bibfnamefont {N.~J.}\ \bibnamefont {Halas}}, \ and\
  \bibinfo {author} {\bibfnamefont {P.}~\bibnamefont {Nordlander}},\
  }\href@noop {} {\bibfield  {journal} {\bibinfo  {journal} {Science}\ }\textbf
  {\bibinfo {volume} {302}},\ \bibinfo {pages} {419} (\bibinfo {year}
  {2003})}\BibitemShut {NoStop}%
\bibitem [{\citenamefont {Liu}\ \emph {et~al.}(2009)\citenamefont {Liu},
  \citenamefont {Langguth}, \citenamefont {Weiss}, \citenamefont {K\"{a}stel},
  \citenamefont {Fleischhauer}, \citenamefont {Pfau},\ and\ \citenamefont
  {Giessen}}]{LiuEtAlNatMat2009}%
  \BibitemOpen
  \bibfield  {author} {\bibinfo {author} {\bibfnamefont {N.}~\bibnamefont
  {Liu}}, \bibinfo {author} {\bibfnamefont {L.}~\bibnamefont {Langguth}},
  \bibinfo {author} {\bibfnamefont {T.}~\bibnamefont {Weiss}}, \bibinfo
  {author} {\bibfnamefont {J.}~\bibnamefont {K\"{a}stel}}, \bibinfo {author}
  {\bibfnamefont {M.}~\bibnamefont {Fleischhauer}}, \bibinfo {author}
  {\bibfnamefont {T.}~\bibnamefont {Pfau}}, \ and\ \bibinfo {author}
  {\bibfnamefont {H.}~\bibnamefont {Giessen}},\ }\href@noop {} {\bibfield
  {journal} {\bibinfo  {journal} {Nat. Mater.}\ }\textbf {\bibinfo {volume}
  {8}},\ \bibinfo {pages} {758} (\bibinfo {year} {2009})}\BibitemShut {NoStop}%
\bibitem [{\citenamefont {Lovera}\ \emph {et~al.}(2013)\citenamefont {Lovera},
  \citenamefont {Gallinet}, \citenamefont {Nordlander},\ and\ \citenamefont
  {Martin}}]{Lovera}%
  \BibitemOpen
  \bibfield  {author} {\bibinfo {author} {\bibfnamefont {A.}~\bibnamefont
  {Lovera}}, \bibinfo {author} {\bibfnamefont {B.}~\bibnamefont {Gallinet}},
  \bibinfo {author} {\bibfnamefont {P.}~\bibnamefont {Nordlander}}, \ and\
  \bibinfo {author} {\bibfnamefont {O.~J.}\ \bibnamefont {Martin}},\ }\href
  {\doibase 10.1021/nn401175j} {\bibfield  {journal} {\bibinfo  {journal} {ACS
  Nano}\ }\textbf {\bibinfo {volume} {7}},\ \bibinfo {pages} {4527} (\bibinfo
  {year} {2013})}\BibitemShut {NoStop}%
\bibitem [{\citenamefont {Fan}\ \emph {et~al.}(2010)\citenamefont {Fan},
  \citenamefont {Wu}, \citenamefont {Bao}, \citenamefont {Bao}, \citenamefont
  {Bardhan}, \citenamefont {Halas}, \citenamefont {Manoharan}, \citenamefont
  {Nordlander}, \citenamefont {Shvets},\ and\ \citenamefont
  {Capasso}}]{FanCapasso}%
  \BibitemOpen
  \bibfield  {author} {\bibinfo {author} {\bibfnamefont {J.~A.}\ \bibnamefont
  {Fan}}, \bibinfo {author} {\bibfnamefont {C.}~\bibnamefont {Wu}}, \bibinfo
  {author} {\bibfnamefont {K.}~\bibnamefont {Bao}}, \bibinfo {author}
  {\bibfnamefont {J.}~\bibnamefont {Bao}}, \bibinfo {author} {\bibfnamefont
  {R.}~\bibnamefont {Bardhan}}, \bibinfo {author} {\bibfnamefont {N.~J.}\
  \bibnamefont {Halas}}, \bibinfo {author} {\bibfnamefont {V.~N.}\ \bibnamefont
  {Manoharan}}, \bibinfo {author} {\bibfnamefont {P.}~\bibnamefont
  {Nordlander}}, \bibinfo {author} {\bibfnamefont {G.}~\bibnamefont {Shvets}},
  \ and\ \bibinfo {author} {\bibfnamefont {F.}~\bibnamefont {Capasso}},\ }\href
  {\doibase 10.1126/science.1187949} {\bibfield  {journal} {\bibinfo  {journal}
  {Science}\ }\textbf {\bibinfo {volume} {328}},\ \bibinfo {pages} {1135}
  (\bibinfo {year} {2010})}\BibitemShut {NoStop}%
\bibitem [{\citenamefont {Hentschel}\ \emph {et~al.}(2011)\citenamefont
  {Hentschel}, \citenamefont {Dregely}, \citenamefont {Vogelgesang},
  \citenamefont {Giessen},\ and\ \citenamefont {Liu}}]{GiessenOligomers}%
  \BibitemOpen
  \bibfield  {author} {\bibinfo {author} {\bibfnamefont {M.}~\bibnamefont
  {Hentschel}}, \bibinfo {author} {\bibfnamefont {D.}~\bibnamefont {Dregely}},
  \bibinfo {author} {\bibfnamefont {R.}~\bibnamefont {Vogelgesang}}, \bibinfo
  {author} {\bibfnamefont {H.}~\bibnamefont {Giessen}}, \ and\ \bibinfo
  {author} {\bibfnamefont {N.}~\bibnamefont {Liu}},\ }\href {\doibase
  10.1021/nn103172t} {\bibfield  {journal} {\bibinfo  {journal} {ACS Nano}\
  }\textbf {\bibinfo {volume} {5}},\ \bibinfo {pages} {2042} (\bibinfo {year}
  {2011})}\BibitemShut {NoStop}%
\bibitem [{\citenamefont {Frimmer}\ \emph {et~al.}(2012)\citenamefont
  {Frimmer}, \citenamefont {Coenen},\ and\ \citenamefont
  {Koenderink}}]{Frimmer}%
  \BibitemOpen
  \bibfield  {author} {\bibinfo {author} {\bibfnamefont {M.}~\bibnamefont
  {Frimmer}}, \bibinfo {author} {\bibfnamefont {T.}~\bibnamefont {Coenen}}, \
  and\ \bibinfo {author} {\bibfnamefont {A.~F.}\ \bibnamefont {Koenderink}},\
  }\href {\doibase {10.1103/PhysRevLett.108.077404}} {\bibfield  {journal}
  {\bibinfo  {journal} {{Phys. Rev. Lett.}}\ }\textbf {\bibinfo {volume}
  {{108}}},\ \bibinfo {pages} {{077404}} (\bibinfo {year}
  {{2012}})}\BibitemShut {NoStop}%
\bibitem [{\citenamefont {Dregely}\ \emph {et~al.}(2011)\citenamefont
  {Dregely}, \citenamefont {Hentschel},\ and\ \citenamefont
  {Giessen}}]{Dregely}%
  \BibitemOpen
  \bibfield  {author} {\bibinfo {author} {\bibfnamefont {D.}~\bibnamefont
  {Dregely}}, \bibinfo {author} {\bibfnamefont {M.}~\bibnamefont {Hentschel}},
  \ and\ \bibinfo {author} {\bibfnamefont {H.}~\bibnamefont {Giessen}},\ }\href
  {\doibase 10.1021/nn202876k} {\bibfield  {journal} {\bibinfo  {journal} {ACS
  Nano}\ }\textbf {\bibinfo {volume} {5}},\ \bibinfo {pages} {8202} (\bibinfo
  {year} {2011})}\BibitemShut {NoStop}%
\bibitem [{\citenamefont {Watson}\ \emph {et~al.}(2016)\citenamefont {Watson},
  \citenamefont {Jenkins}, \citenamefont {Ruostekoski}, \citenamefont
  {Fedotov},\ and\ \citenamefont {Zheludev}}]{Watson2016}%
  \BibitemOpen
  \bibfield  {author} {\bibinfo {author} {\bibfnamefont {D.~W.}\ \bibnamefont
  {Watson}}, \bibinfo {author} {\bibfnamefont {S.~D.}\ \bibnamefont {Jenkins}},
  \bibinfo {author} {\bibfnamefont {J.}~\bibnamefont {Ruostekoski}}, \bibinfo
  {author} {\bibfnamefont {V.~A.}\ \bibnamefont {Fedotov}}, \ and\ \bibinfo
  {author} {\bibfnamefont {N.~I.}\ \bibnamefont {Zheludev}},\ }\href@noop {}
  {\bibfield  {journal} {\bibinfo  {journal} {Phys. Rev. B}\ }\textbf {\bibinfo
  {volume} {93}},\ \bibinfo {pages} {125420} (\bibinfo {year}
  {2016})}\BibitemShut {NoStop}%
\bibitem [{\citenamefont {Hopkins}\ \emph {et~al.}(2013)\citenamefont
  {Hopkins}, \citenamefont {Poddubny}, \citenamefont {Miroshnichenko},\ and\
  \citenamefont {Kivshar}}]{Hopkins13}%
  \BibitemOpen
  \bibfield  {author} {\bibinfo {author} {\bibfnamefont {B.}~\bibnamefont
  {Hopkins}}, \bibinfo {author} {\bibfnamefont {A.~N.}\ \bibnamefont
  {Poddubny}}, \bibinfo {author} {\bibfnamefont {A.~E.}\ \bibnamefont
  {Miroshnichenko}}, \ and\ \bibinfo {author} {\bibfnamefont {Y.~S.}\
  \bibnamefont {Kivshar}},\ }\href {\doibase 10.1103/PhysRevA.88.053819}
  {\bibfield  {journal} {\bibinfo  {journal} {Phys. Rev. A}\ }\textbf {\bibinfo
  {volume} {88}},\ \bibinfo {pages} {053819} (\bibinfo {year}
  {2013})}\BibitemShut {NoStop}%
\bibitem [{\citenamefont {Forestiere}\ \emph {et~al.}(2013)\citenamefont
  {Forestiere}, \citenamefont {Dal~Negro},\ and\ \citenamefont
  {Miano}}]{Forestiere13}%
  \BibitemOpen
  \bibfield  {author} {\bibinfo {author} {\bibfnamefont {C.}~\bibnamefont
  {Forestiere}}, \bibinfo {author} {\bibfnamefont {L.}~\bibnamefont
  {Dal~Negro}}, \ and\ \bibinfo {author} {\bibfnamefont {G.}~\bibnamefont
  {Miano}},\ }\href {\doibase 10.1103/PhysRevB.88.155411} {\bibfield  {journal}
  {\bibinfo  {journal} {Phys. Rev. B}\ }\textbf {\bibinfo {volume} {88}},\
  \bibinfo {pages} {155411} (\bibinfo {year} {2013})}\BibitemShut {NoStop}%
\bibitem [{\citenamefont {Papasimakis}\ \emph {et~al.}(2009)\citenamefont
  {Papasimakis}, \citenamefont {Fedotov}, \citenamefont {Fu}, \citenamefont
  {Tsai},\ and\ \citenamefont {Zheludev}}]{papasimakis2009}%
  \BibitemOpen
  \bibfield  {author} {\bibinfo {author} {\bibfnamefont {N.}~\bibnamefont
  {Papasimakis}}, \bibinfo {author} {\bibfnamefont {V.~A.}\ \bibnamefont
  {Fedotov}}, \bibinfo {author} {\bibfnamefont {Y.~H.}\ \bibnamefont {Fu}},
  \bibinfo {author} {\bibfnamefont {D.~P.}\ \bibnamefont {Tsai}}, \ and\
  \bibinfo {author} {\bibfnamefont {N.~I.}\ \bibnamefont {Zheludev}},\
  }\href@noop {} {\bibfield  {journal} {\bibinfo  {journal} {Phys. Rev. B}\
  }\textbf {\bibinfo {volume} {80}},\ \bibinfo {pages} {041102(R)} (\bibinfo
  {year} {2009})}\BibitemShut {NoStop}%
\bibitem [{\citenamefont {Fedotov}\ \emph {et~al.}(2010)\citenamefont
  {Fedotov}, \citenamefont {Papasimakis}, \citenamefont {Plum}, \citenamefont
  {Bitzer}, \citenamefont {Walther}, \citenamefont {Kuo}, \citenamefont
  {Tsai},\ and\ \citenamefont {Zheludev}}]{FedotovEtAlPRL2010}%
  \BibitemOpen
  \bibfield  {author} {\bibinfo {author} {\bibfnamefont {V.~A.}\ \bibnamefont
  {Fedotov}}, \bibinfo {author} {\bibfnamefont {N.}~\bibnamefont
  {Papasimakis}}, \bibinfo {author} {\bibfnamefont {E.}~\bibnamefont {Plum}},
  \bibinfo {author} {\bibfnamefont {A.}~\bibnamefont {Bitzer}}, \bibinfo
  {author} {\bibfnamefont {M.}~\bibnamefont {Walther}}, \bibinfo {author}
  {\bibfnamefont {P.}~\bibnamefont {Kuo}}, \bibinfo {author} {\bibfnamefont
  {D.~P.}\ \bibnamefont {Tsai}}, \ and\ \bibinfo {author} {\bibfnamefont
  {N.~I.}\ \bibnamefont {Zheludev}},\ }\href@noop {} {\bibfield  {journal}
  {\bibinfo  {journal} {Phys. Rev. Lett.}\ }\textbf {\bibinfo {volume} {104}},\
  \bibinfo {pages} {223901} (\bibinfo {year} {2010})}\BibitemShut {NoStop}%
\bibitem [{\citenamefont {Holstein}(1947)}]{Holstein}%
  \BibitemOpen
  \bibfield  {author} {\bibinfo {author} {\bibfnamefont {T.}~\bibnamefont
  {Holstein}},\ }\href {\doibase 10.1103/PhysRev.72.1212} {\bibfield  {journal}
  {\bibinfo  {journal} {Phys. Rev.}\ }\textbf {\bibinfo {volume} {72}},\
  \bibinfo {pages} {1212} (\bibinfo {year} {1947})}\BibitemShut {NoStop}%
\bibitem [{\citenamefont {Cummings}(1986)}]{Cummings86}%
  \BibitemOpen
  \bibfield  {author} {\bibinfo {author} {\bibfnamefont {F.~W.}\ \bibnamefont
  {Cummings}},\ }\href {\doibase 10.1103/PhysRevA.33.1683} {\bibfield
  {journal} {\bibinfo  {journal} {Phys. Rev. A}\ }\textbf {\bibinfo {volume}
  {33}},\ \bibinfo {pages} {1683} (\bibinfo {year} {1986})}\BibitemShut
  {NoStop}%
\bibitem [{\citenamefont {Zheludev}\ \emph {et~al.}(2008)\citenamefont
  {Zheludev}, \citenamefont {Prosvirnin}, \citenamefont {Papasimakis},\ and\
  \citenamefont {Fedotov}}]{ZheludevEtAlNatPhot2008}%
  \BibitemOpen
  \bibfield  {author} {\bibinfo {author} {\bibfnamefont {N.~I.}\ \bibnamefont
  {Zheludev}}, \bibinfo {author} {\bibfnamefont {S.~L.}\ \bibnamefont
  {Prosvirnin}}, \bibinfo {author} {\bibfnamefont {N.}~\bibnamefont
  {Papasimakis}}, \ and\ \bibinfo {author} {\bibfnamefont {V.~A.}\ \bibnamefont
  {Fedotov}},\ }\href@noop {} {\bibfield  {journal} {\bibinfo  {journal}
  {Nature Photonics}\ }\textbf {\bibinfo {volume} {351}} (\bibinfo {year}
  {2008})}\BibitemShut {NoStop}%
\bibitem [{\citenamefont {Jenkins}\ and\ \citenamefont
  {Ruostekoski}(2013)}]{CAIT}%
  \BibitemOpen
  \bibfield  {author} {\bibinfo {author} {\bibfnamefont {S.~D.}\ \bibnamefont
  {Jenkins}}\ and\ \bibinfo {author} {\bibfnamefont {J.}~\bibnamefont
  {Ruostekoski}},\ }\href@noop {} {\bibfield  {journal} {\bibinfo  {journal}
  {Phys. Rev. Lett.}\ }\textbf {\bibinfo {volume} {111}},\ \bibinfo {pages}
  {147401} (\bibinfo {year} {2013})}\BibitemShut {NoStop}%
\bibitem [{\citenamefont {Lemoult}\ \emph {et~al.}(2010)\citenamefont
  {Lemoult}, \citenamefont {Lerosey}, \citenamefont {{de Rosny}},\ and\
  \citenamefont {Fink}}]{LemoultPRL10}%
  \BibitemOpen
  \bibfield  {author} {\bibinfo {author} {\bibfnamefont {F.}~\bibnamefont
  {Lemoult}}, \bibinfo {author} {\bibfnamefont {G.}~\bibnamefont {Lerosey}},
  \bibinfo {author} {\bibfnamefont {J.}~\bibnamefont {{de Rosny}}}, \ and\
  \bibinfo {author} {\bibfnamefont {M.}~\bibnamefont {Fink}},\ }\href {\doibase
  10.1103/PhysRevLett.104.203901} {\bibfield  {journal} {\bibinfo  {journal}
  {Phys. Rev. Lett.}\ }\textbf {\bibinfo {volume} {104}},\ \bibinfo {pages}
  {203901} (\bibinfo {year} {2010})}\BibitemShut {NoStop}%
\bibitem [{\citenamefont {Linden}\ \emph {et~al.}(2012)\citenamefont {Linden},
  \citenamefont {Niesler}, \citenamefont {F\"orstner}, \citenamefont {Grynko},
  \citenamefont {Meier},\ and\ \citenamefont {Wegener}}]{Linden12}%
  \BibitemOpen
  \bibfield  {author} {\bibinfo {author} {\bibfnamefont {S.}~\bibnamefont
  {Linden}}, \bibinfo {author} {\bibfnamefont {F.~B.~P.}\ \bibnamefont
  {Niesler}}, \bibinfo {author} {\bibfnamefont {J.}~\bibnamefont {F\"orstner}},
  \bibinfo {author} {\bibfnamefont {Y.}~\bibnamefont {Grynko}}, \bibinfo
  {author} {\bibfnamefont {T.}~\bibnamefont {Meier}}, \ and\ \bibinfo {author}
  {\bibfnamefont {M.}~\bibnamefont {Wegener}},\ }\href {\doibase
  10.1103/PhysRevLett.109.015502} {\bibfield  {journal} {\bibinfo  {journal}
  {Phys. Rev. Lett.}\ }\textbf {\bibinfo {volume} {109}},\ \bibinfo {pages}
  {015502} (\bibinfo {year} {2012})}\BibitemShut {NoStop}%
\bibitem [{\citenamefont {Fedotov}\ \emph {et~al.}(2007)\citenamefont
  {Fedotov}, \citenamefont {Rose}, \citenamefont {Prosvirnin}, \citenamefont
  {Papasimakis},\ and\ \citenamefont {Zheludev}}]{FedotovEtAlPRL2007}%
  \BibitemOpen
  \bibfield  {author} {\bibinfo {author} {\bibfnamefont {V.~A.}\ \bibnamefont
  {Fedotov}}, \bibinfo {author} {\bibfnamefont {M.}~\bibnamefont {Rose}},
  \bibinfo {author} {\bibfnamefont {S.~L.}\ \bibnamefont {Prosvirnin}},
  \bibinfo {author} {\bibfnamefont {N.}~\bibnamefont {Papasimakis}}, \ and\
  \bibinfo {author} {\bibfnamefont {N.~I.}\ \bibnamefont {Zheludev}},\
  }\href@noop {} {\bibfield  {journal} {\bibinfo  {journal} {Phys. Rev. Lett.}\
  }\textbf {\bibinfo {volume} {99}},\ \bibinfo {pages} {147401} (\bibinfo
  {year} {2007})}\BibitemShut {NoStop}%
\bibitem [{\citenamefont {Ishimaru}(1978)}]{Ishimaru1978}%
  \BibitemOpen
  \bibfield  {author} {\bibinfo {author} {\bibfnamefont {A.}~\bibnamefont
  {Ishimaru}},\ }\href@noop {} {\emph {\bibinfo {title} {Wave Propagation and
  Scattering in Random Media: Multiple Scattering, Turbulence, Rough Surfaces,
  and Remote-Sensing}}},\ Vol.~\bibinfo {volume} {2}\ (\bibinfo  {publisher}
  {Academic Press},\ \bibinfo {address} {St. Louis, Missouri},\ \bibinfo {year}
  {1978})\BibitemShut {NoStop}%
\bibitem [{\citenamefont {Morice}\ \emph {et~al.}(1995)\citenamefont {Morice},
  \citenamefont {Castin},\ and\ \citenamefont {Dalibard}}]{Morice1995a}%
  \BibitemOpen
  \bibfield  {author} {\bibinfo {author} {\bibfnamefont {O.}~\bibnamefont
  {Morice}}, \bibinfo {author} {\bibfnamefont {Y.}~\bibnamefont {Castin}}, \
  and\ \bibinfo {author} {\bibfnamefont {J.}~\bibnamefont {Dalibard}},\
  }\href@noop {} {\bibfield  {journal} {\bibinfo  {journal} {Phys. Rev. A}\
  }\textbf {\bibinfo {volume} {51}},\ \bibinfo {pages} {3896} (\bibinfo {year}
  {1995})}\BibitemShut {NoStop}%
\bibitem [{\citenamefont {Wiersma}\ \emph {et~al.}(1995)\citenamefont
  {Wiersma}, \citenamefont {van Albada}, \citenamefont {van Tiggelen},\ and\
  \citenamefont {Lagendijk}}]{Wiersma_recurrent}%
  \BibitemOpen
  \bibfield  {author} {\bibinfo {author} {\bibfnamefont {D.~S.}\ \bibnamefont
  {Wiersma}}, \bibinfo {author} {\bibfnamefont {M.~P.}\ \bibnamefont {van
  Albada}}, \bibinfo {author} {\bibfnamefont {B.~A.}\ \bibnamefont {van
  Tiggelen}}, \ and\ \bibinfo {author} {\bibfnamefont {A.}~\bibnamefont
  {Lagendijk}},\ }\href {\doibase 10.1103/PhysRevLett.74.4193} {\bibfield
  {journal} {\bibinfo  {journal} {Phys. Rev. Lett.}\ }\textbf {\bibinfo
  {volume} {74}},\ \bibinfo {pages} {4193} (\bibinfo {year}
  {1995})}\BibitemShut {NoStop}%
\bibitem [{\citenamefont {Ruostekoski}\ and\ \citenamefont
  {Javanainen}(1997)}]{Ruostekoski1997a}%
  \BibitemOpen
  \bibfield  {author} {\bibinfo {author} {\bibfnamefont {J.}~\bibnamefont
  {Ruostekoski}}\ and\ \bibinfo {author} {\bibfnamefont {J.}~\bibnamefont
  {Javanainen}},\ }\href@noop {} {\bibfield  {journal} {\bibinfo  {journal}
  {Phys. Rev. A}\ }\textbf {\bibinfo {volume} {55}},\ \bibinfo {pages} {513}
  (\bibinfo {year} {1997})}\BibitemShut {NoStop}%
\bibitem [{\citenamefont {Javanainen}\ \emph {et~al.}(2014)\citenamefont
  {Javanainen}, \citenamefont {Ruostekoski}, \citenamefont {Li},\ and\
  \citenamefont {Yoo}}]{Javanainen2014a}%
  \BibitemOpen
  \bibfield  {author} {\bibinfo {author} {\bibfnamefont {J.}~\bibnamefont
  {Javanainen}}, \bibinfo {author} {\bibfnamefont {J.}~\bibnamefont
  {Ruostekoski}}, \bibinfo {author} {\bibfnamefont {Y.}~\bibnamefont {Li}}, \
  and\ \bibinfo {author} {\bibfnamefont {S.-M.}\ \bibnamefont {Yoo}},\ }\href
  {\doibase 10.1103/PhysRevLett.112.113603} {\bibfield  {journal} {\bibinfo
  {journal} {Phys. Rev. Lett.}\ }\textbf {\bibinfo {volume} {112}},\ \bibinfo
  {pages} {113603} (\bibinfo {year} {2014})}\BibitemShut {NoStop}%
\bibitem [{\citenamefont {Javanainen}\ and\ \citenamefont
  {Ruostekoski}(2016)}]{JavanainenMFT}%
  \BibitemOpen
  \bibfield  {author} {\bibinfo {author} {\bibfnamefont {J.}~\bibnamefont
  {Javanainen}}\ and\ \bibinfo {author} {\bibfnamefont {J.}~\bibnamefont
  {Ruostekoski}},\ }\href {\doibase 10.1364/OE.24.000993} {\bibfield  {journal}
  {\bibinfo  {journal} {Opt. Express}\ }\textbf {\bibinfo {volume} {24}},\
  \bibinfo {pages} {993} (\bibinfo {year} {2016})}\BibitemShut {NoStop}%
\bibitem [{\citenamefont {Jenkins}\ and\ \citenamefont
  {Ruostekoski}(2012{\natexlab{a}})}]{JenkinsLongPRB}%
  \BibitemOpen
  \bibfield  {author} {\bibinfo {author} {\bibfnamefont {S.~D.}\ \bibnamefont
  {Jenkins}}\ and\ \bibinfo {author} {\bibfnamefont {J.}~\bibnamefont
  {Ruostekoski}},\ }\href@noop {} {\bibfield  {journal} {\bibinfo  {journal}
  {Phys. Rev. B}\ }\textbf {\bibinfo {volume} {86}},\ \bibinfo {pages} {085116}
  (\bibinfo {year} {2012}{\natexlab{a}})}\BibitemShut {NoStop}%
\bibitem [{SOM()}]{SOM}%
  \BibitemOpen
  \href@noop {} {}\bibinfo {note} {See Supplemental Material for additional
  details regarding technical background information.}\BibitemShut {Stop}%
\bibitem [{\citenamefont {Savo}\ \emph {et~al.}(2012)\citenamefont {Savo},
  \citenamefont {Papasimakis},\ and\ \citenamefont
  {Zheludev}}]{SavoEtAlPRB2012}%
  \BibitemOpen
  \bibfield  {author} {\bibinfo {author} {\bibfnamefont {S.}~\bibnamefont
  {Savo}}, \bibinfo {author} {\bibfnamefont {N.}~\bibnamefont {Papasimakis}}, \
  and\ \bibinfo {author} {\bibfnamefont {N.~I.}\ \bibnamefont {Zheludev}},\
  }\href@noop {} {\bibfield  {journal} {\bibinfo  {journal} {Phys. Rev. B}\
  }\textbf {\bibinfo {volume} {85}},\ \bibinfo {pages} {121104(R)} (\bibinfo
  {year} {2012})}\BibitemShut {NoStop}%
\bibitem [{\citenamefont {Kuwata}\ \emph {et~al.}(2003)\citenamefont {Kuwata},
  \citenamefont {Tamaru}, \citenamefont {Esumi},\ and\ \citenamefont
  {Miyano}}]{kuwate}%
  \BibitemOpen
  \bibfield  {author} {\bibinfo {author} {\bibfnamefont {H.}~\bibnamefont
  {Kuwata}}, \bibinfo {author} {\bibfnamefont {H.}~\bibnamefont {Tamaru}},
  \bibinfo {author} {\bibfnamefont {K.}~\bibnamefont {Esumi}}, \ and\ \bibinfo
  {author} {\bibfnamefont {K.}~\bibnamefont {Miyano}},\ }\href@noop {}
  {\bibfield  {journal} {\bibinfo  {journal} {Applied Physics Letters}\
  }\textbf {\bibinfo {volume} {83}},\ \bibinfo {pages} {4625} (\bibinfo {year}
  {2003})}\BibitemShut {NoStop}%
\bibitem [{\citenamefont {Guerin}\ \emph {et~al.}(2016)\citenamefont {Guerin},
  \citenamefont {Ara\'ujo},\ and\ \citenamefont {Kaiser}}]{Guerin_subr16}%
  \BibitemOpen
  \bibfield  {author} {\bibinfo {author} {\bibfnamefont {W.}~\bibnamefont
  {Guerin}}, \bibinfo {author} {\bibfnamefont {M.~O.}\ \bibnamefont
  {Ara\'ujo}}, \ and\ \bibinfo {author} {\bibfnamefont {R.}~\bibnamefont
  {Kaiser}},\ }\href {\doibase 10.1103/PhysRevLett.116.083601} {\bibfield
  {journal} {\bibinfo  {journal} {Phys. Rev. Lett.}\ }\textbf {\bibinfo
  {volume} {116}},\ \bibinfo {pages} {083601} (\bibinfo {year}
  {2016})}\BibitemShut {NoStop}%
\bibitem [{\citenamefont {Jenkins}\ and\ \citenamefont
  {Ruostekoski}(2012{\natexlab{b}})}]{Jenkins2012a}%
  \BibitemOpen
  \bibfield  {author} {\bibinfo {author} {\bibfnamefont {S.~D.}\ \bibnamefont
  {Jenkins}}\ and\ \bibinfo {author} {\bibfnamefont {J.}~\bibnamefont
  {Ruostekoski}},\ }\href@noop {} {\bibfield  {journal} {\bibinfo  {journal}
  {Phys. Rev. A}\ }\textbf {\bibinfo {volume} {86}},\ \bibinfo {pages}
  {031602(R)} (\bibinfo {year} {2012}{\natexlab{b}})}\BibitemShut {NoStop}%
\bibitem [{\citenamefont {Bettles}\ \emph {et~al.}(2015)\citenamefont
  {Bettles}, \citenamefont {Gardiner},\ and\ \citenamefont
  {Adams}}]{Bettles2014}%
  \BibitemOpen
  \bibfield  {author} {\bibinfo {author} {\bibfnamefont {R.~J.}\ \bibnamefont
  {Bettles}}, \bibinfo {author} {\bibfnamefont {S.~A.}\ \bibnamefont
  {Gardiner}}, \ and\ \bibinfo {author} {\bibfnamefont {C.~S.}\ \bibnamefont
  {Adams}},\ }\href {\doibase 10.1103/PhysRevA.92.063822} {\bibfield  {journal}
  {\bibinfo  {journal} {Phys. Rev. A}\ }\textbf {\bibinfo {volume} {92}},\
  \bibinfo {pages} {063822} (\bibinfo {year} {2015})}\BibitemShut {NoStop}%
\bibitem [{\citenamefont {Lee}\ \emph {et~al.}(2016)\citenamefont {Lee},
  \citenamefont {Jenkins},\ and\ \citenamefont {Ruostekoski}}]{Lee16}%
  \BibitemOpen
  \bibfield  {author} {\bibinfo {author} {\bibfnamefont {M.~D.}\ \bibnamefont
  {Lee}}, \bibinfo {author} {\bibfnamefont {S.~D.}\ \bibnamefont {Jenkins}}, \
  and\ \bibinfo {author} {\bibfnamefont {J.}~\bibnamefont {Ruostekoski}},\
  }\href {\doibase 10.1103/PhysRevA.93.063803} {\bibfield  {journal} {\bibinfo
  {journal} {Phys. Rev. A}\ }\textbf {\bibinfo {volume} {93}},\ \bibinfo
  {pages} {063803} (\bibinfo {year} {2016})}\BibitemShut {NoStop}%
\bibitem [{\citenamefont {Jennewein}\ \emph {et~al.}(2016)\citenamefont
  {Jennewein}, \citenamefont {Besbes}, \citenamefont {Schilder}, \citenamefont
  {Jenkins}, \citenamefont {Sauvan}, \citenamefont {Ruostekoski}, \citenamefont
  {Greffet}, \citenamefont {Sortais},\ and\ \citenamefont
  {Browaeys}}]{Jennewein_trans}%
  \BibitemOpen
  \bibfield  {author} {\bibinfo {author} {\bibfnamefont {S.}~\bibnamefont
  {Jennewein}}, \bibinfo {author} {\bibfnamefont {M.}~\bibnamefont {Besbes}},
  \bibinfo {author} {\bibfnamefont {N.~J.}\ \bibnamefont {Schilder}}, \bibinfo
  {author} {\bibfnamefont {S.~D.}\ \bibnamefont {Jenkins}}, \bibinfo {author}
  {\bibfnamefont {C.}~\bibnamefont {Sauvan}}, \bibinfo {author} {\bibfnamefont
  {J.}~\bibnamefont {Ruostekoski}}, \bibinfo {author} {\bibfnamefont {J.-J.}\
  \bibnamefont {Greffet}}, \bibinfo {author} {\bibfnamefont {Y.~R.~P.}\
  \bibnamefont {Sortais}}, \ and\ \bibinfo {author} {\bibfnamefont
  {A.}~\bibnamefont {Browaeys}},\ }\href {\doibase
  10.1103/PhysRevLett.116.233601} {\bibfield  {journal} {\bibinfo  {journal}
  {Phys. Rev. Lett.}\ }\textbf {\bibinfo {volume} {116}},\ \bibinfo {pages}
  {233601} (\bibinfo {year} {2016})}\BibitemShut {NoStop}%
 \bibitem [{\citenamefont {Jackson}(1999)}]{Jackson}%
  \BibitemOpen
  \bibfield  {author} {\bibinfo {author} {\bibfnamefont {J.~D.}\ \bibnamefont
  {Jackson}},\ }\href@noop {} {\emph {\bibinfo {title} {Classical
  Electrodynamics}}},\ \bibinfo {edition} {3rd}\ ed.\ (\bibinfo  {publisher}
  {Wiley, New York},\ \bibinfo {year} {1999})\BibitemShut {NoStop}%
\end{thebibliography}
\end{document}